\documentclass[page-classic]{epl2} 
\usepackage{graphicx}
\usepackage{amsmath}
\usepackage[bbgreekl]{mathbbol}
\usepackage{amssymb}
\usepackage{bm}
\usepackage{subfigure}
\usepackage{slashed}
\usepackage{hyperref}

\def\0{\mbox{\boldmath$\displaystyle\mathbf{0}$}}

\title{Magnetic field creation by solar mass neutrino jets}
\shorttitle{Title} 

\author{Dharam Vir Ahluwalia\inst{1} \and Cheng-Yang Lee\inst{2}}

\institute{                    
  \inst{1} Centre for the Studies of the Glass Bead Game\\
Chaugon, Bir, Himachal Pradesh 176 077, India\\

  \inst{2} Manipal Centre for Natural Sciences, Centre of Excellence,\\ Manipal Academy of Higher Education\\ Dr TMA Pai Planetarium Building, Manipal 576104, Karnataka, India
}

\abstract{Parity violation and its effects for neutrinos in astrophysical contexts have been considered earlier in pioneering papers of Hawking and Vilenkin. But because even the largest magnetic moments predicted by physics beyond the Standard Model are some twelve orders of magnitude smaller than the Bohr magneton, their implications for magnetic field generation and neutrino oscillations are generally considered insignificant. Here we show that since in astrophysical scenarios a huge number of neutrinos may be emitted, the smallness of the magnetic moment, when coupled with parity violation, is compensated by the sheer number of neutrinos. The merger of neutron stars would leave behind a short pulse of electromagnetic synchrotron radiation even if the neutrino jet in the merger points away from the neutrino detectors. We show that the magnetic field can be as large as $10^6\,\mbox{Gauss}$ and comment on the possibility of direct detection. Observation of such a pulse would lend strong support for neutrino magnetic moments and resolve the missing neutrino problem in neutron star mergers. }

\def\mubf{\mbox{\boldmath$\displaystyle\boldsymbol{\mathfrak{\mu}}$}}

\def\mbf{\mbox{\boldmath$\displaystyle\boldsymbol{\mathfrak{m}}$}}

\def\Bbf{\mbox{\boldmath$\displaystyle\boldsymbol{\mathfrak{B}}$}}

\def\Ebf{\mbox{\boldmath$\displaystyle\boldsymbol{\mathfrak{E}}$}}

\begin{document}

\maketitle

Significance of neutrinos and consequence of parity violation in astrophysical processes appeared in early works of Hawking and Vilenkin~\cite{Hawking:1974sw,PhysRevLett.41.1575}, and more recently in the context of pulsar kicks in~\cite{Sagert:2007as,Kisslinger:2009rv}, while effects of magnetic fields on neutrino oscillations were brought to attention in~\cite{Ahluwalia:2017rsu}. 
The Dirac and Majorana magnetic moments that are nearly ten orders of magnitude larger than that in the minimal extension of the standard model were discussed in~\cite{Babu:1987be}. Here, we report a new effect in the same series and in particular discuss the possibility of confirming or ruling out models such as these suggested in~\cite{Babu:1987be}. 

The helicity of an  ultra-relativistic neutrino, if it is produced in an electroweak process, points in a direction
opposite  to its direction of motion. 
This fact has the consequence  that for an astrophysical  neutrino jet, all the helicities are pointed in essentially the same direction, giving
rise to a magnetic moment that depends on the product of a
very tiny neutrino magnetic moment with a huge number density of
neutrinos (perhaps a solar mass or so of them moving co-linearly). The resulting magnetic field gives
rise to a synchrotron radiation if the jet passes through a plasma. 

More concretely, we couple the following three observations: (a) In many astrophysical explosions/processes a solar mass of neutrinos may be emitted on a time scale of a few seconds (these, in a broad brush, may be co-linear, or radial -- here, we focus on the former),  (b) A massive neutrino is inevitably endowed with a tiny magnetic moment, (c) A neutrino emitted in parity violating electroweak  processes has its helicity, and hence its magnetic moment, in a direction either parallel or antiparallel -- depending on whether we are focusing on an antiparticle or a particle -- to the direction of its propagation. Therefore, under the assumption of strong CP violation, a  huge number of neutrinos or anti-neutrinos can compensate a tiny magnetic moment to generate large magnetic fields.





Since neutrinos of the standard model of high energy physics are massless,
massive neutrinos suggested by the observed neutrino oscillations constitute  an evidence for the physics beyond the standard model and have important ramifications for astrophysics and cosmology. Apart from the mentioned phenomenon of flavour oscillations, another important consequence, owing to their masses is that the underlying mass eigenstates acquire  non-zero magnetic  moments. To the leading order in $\left(m_i/m_W\right)^2$ the magnetic moments for the Dirac neutrinos in the minimal extended SM are given by~\cite{PhysRevLett.45.963,Shrock:1982sc}

\begin{eqnarray}
\mubf_{i}&=& - \,\frac{3 e G_{F} m_{i}}{8 \sqrt{2} \pi^{2}} \, \widehat{\textbf{p}}\nonumber\\
&=& - \,1.85\times 10^{-27}\left(\frac{m_{i}}{\mbox{eV}}\right)\left(\frac{\mbox{eV}}{\mbox{Gauss}}\right)\, \widehat{\textbf{p}}\\
&=&-3.2\times 10^{-19}\left(\frac{m_{i}}{\mbox{eV}}\right)\mubf_{B}
\end{eqnarray}
where $m_{i}$, $i=1,2,3$,  is the mass of the \emph{ith} mass eigenstate, $\mu_{B}$ is the Bohr magneton, $G_{F}$ is the Fermi constant, and $ \widehat{\textbf{p}}$ is a unit vector in the direction of neutrino propagation. In the above expression, $e > 0$. In contrast, for Majorana neutrinos $\mubf_i$ identically vanish.
The magnetic moments of the flavour eigenstates are then simply
\begin{equation}
\mubf_\alpha = \sum_{i=1,2,3} U_{\alpha i}^\ast U_{\alpha i} \mubf_i,\qquad \mbox{(no sum on $\alpha$)}
\end{equation}
with $\alpha = e,\mu,\tau$. Here, $U_{\alpha i}$ is a unitary matrix that encodes 
neutrino mixing.

We model an astrophysical neutrino jet
 as a cylinder of length $\ell \gg r$, with $\ell$ being the temporal size of the jet and $r$ as its radius, with uniform magnetization antiparallel to the co-linearly moving neutrinos. It is endowed with a magnetization 
 \begin{equation}
 \mbf_\alpha = n_\alpha\, \mubf_{\alpha},\qquad \mbox{(no sum on $\alpha$)}
 \end{equation} 
 where $n_\alpha $ is the neutrino number density and is assumed to be uniform and constant throughout the jet. This immediately yields the following result for the magnetic field associated with flavour $\alpha$
 \begin{equation}
 \Bbf_\alpha = -\, 4\times 10^{-38}\left(\frac{n_\alpha}{\mbox{cm}^{-3}}\right)
\left(\frac{\sum_{i=1,2,3} {U^\ast_{\alpha i}} U_{\alpha i} m_i}{\mbox{eV}}\right) \widehat{\textbf{p}} ~\mbox{Gauss} \label{eq:b}
 \end{equation}
 for the inside of the jet, and zero for the outside. Ignoring a LSND-suggested mass eigenstate of the order of an eV, a reasonable estimate for the neutrino mass is
\begin{equation}
\sum_{i=1,2,3} U_{\alpha i}^\ast U_{\alpha i} m_i\sim 10^{-3}\,\mbox{eV}. \label{eq:mi}
\end{equation}
On the other hand, a LSND-suggested mass eigenstate of the order of an eV would soften this constraint by three orders of magnitude.


As the neutrinos propagate, flavor oscillations induce the following time dependence on the magnetic field
\begin{equation}
\Bbf_{\alpha}(t)=n_{\alpha}\sum_{\beta}P_{\alpha\rightarrow\beta}(t)\mubf_{\alpha}
\end{equation}
where $P_{\alpha\rightarrow \beta}(t)$ is the oscillation probability from flavour $\alpha$ to $\beta$. For relativistic neutrinos, we have
\begin{equation}
P_{\alpha\rightarrow \beta}(t)=\sum_{i,j}U^{*}_{\alpha i}U_{\beta i}U_{\alpha j}U^{*}_{\beta j}\exp\left(-\frac{i\Delta m^{2}_{ij}t}{2E_{\nu}}\right).
\end{equation}
Summing over all flavors, the total magnetic field of the jet is 
\begin{equation}
\Bbf(t)=\sum_{\alpha,\beta}P_{\alpha\rightarrow\beta}(t)n_{\alpha}\mubf_{\beta}
\end{equation}

The time dependence of the magnetic field induces an electric field $\Ebf(t)$ circling the jet. The magnitude of $\Ebf(t)$ is given by
\begin{eqnarray}
|\boldsymbol{\mathfrak{E}}(t)|&=&\left(\frac{r}{2}\right)\sum_{\alpha,\beta}\frac{dP_{\alpha\rightarrow \beta}(t)}{dt}
n_{\alpha}|\mubf_{\beta}| \nonumber\\
&=&\sum_{\alpha,\beta}\sum_{i,j}U^{*}_{\alpha i}U_{\beta i}U_{\alpha j}
U^{*}_{\beta j}\left(-\frac{ir\Delta m^{2}_{ij}}{4E_{\nu}}\right)
\exp\left(-\frac{i\Delta m^{2}_{ij}t}{2E_{\nu}}\right)n_{\alpha}|\mubf_{\beta}|.
\label{eq:efield}
\end{eqnarray}

As the jet passes through a medium consisting of ultra-relativistic electrons, the electrons with non-vanishing velocity components perpendicular to the magnetic field will emit synchrotron radiation. Since the electrons are relativistic, the dominant contribution to the synchrotron radiation comes from the perpendicular component of the acceleration relative to the velocities of the electrons. In other words, we may consider the electrons to be instantaneously traveling along a circular path. 
The electric field only affect the acceleration parallel to the velocities so its contribution to radiation emission is negligible. Here, we take the relativistic electron number density to be $N(\gamma)d\gamma=C\gamma^{-p}d\gamma$ for $\gamma_{1}<\gamma<\gamma_{2}$. For simplicity, we shall assume that the number density distribution holds for a sufficiently wide energy range to warrant the choice $\gamma_{1}=1$ and $\gamma_{2}\rightarrow\infty$. Therefore, the total power emitted per frequency per unit volume in the natural units ($c=\hbar=1$) is given by~\cite{Rybicki:2004hfl}
\begin{equation}
P_{\tiny{\mbox{tot}}}(\omega,t)=\frac{\sqrt{3}e^{3}C|\boldsymbol{\mathfrak{B}}(t)|\sin\delta}{2\pi m_{e}(p+1)}\Gamma\left(\frac{p}{4}+\frac{19}{12}\right)
\Gamma\left(\frac{p}{4}-\frac{1}{12}\right)
\left[\frac{m_{e}\omega}{3e|\boldsymbol{\mathfrak{B}}(t)|\sin\delta}\right]^{-(p-1)/2}.\label{eq:power}
\end{equation}

The time dependence of $|\boldsymbol{\mathfrak{B}}(t)|$ induces an oscillation in $P_{\tiny{\mbox{tot}}}(\omega,t)$. For this effect to be observable, the length of the jet must be at least comparable to the oscillation length $L^{\tiny{\mbox{osc}}}_{ij}=4\pi E_{\nu}/\Delta m^{2}_{ij}$. For neutrinos produced in supernovae, their average energies are $E_{\nu}\sim 10\,\mbox{MeV}$, which corresponds to $L^{\tiny{\mbox{osc}}}_{23}\sim 10^{3}\,\mbox{m}$ and $L^{\tiny{\mbox{osc}}}_{12}\sim 10^{5}\,\mbox{m}$ while the length of the neutrino jet corresponds to the time duration in which they are emitted. The latter is of the order of a few seconds and is much larger than $L^{\tiny{\mbox{osc}}}_{ij}$ so the condition for an oscillating $P_{\tiny{\mbox{tot}}}(\omega,t)$ is satisfied. 

Assuming that the volume of the bulk occupied by the electrons is greater than the volume of the jet $V_{\tiny{\mbox{jet}}}$, the expected spectral flux density is of the order
\begin{equation}
\mathcal{F}(\omega,t)\sim P_{\tiny{\mbox{tot}}}
(\omega,t)[V_{\tiny{\mbox{jet}}}\Gamma^{2}_{\tiny{\mbox{bulk}}}/(\pi D^{2})]\label{eq:flux}
\end{equation}
 where $D$ is the distance between the source and the detector and $\Gamma_{\tiny{\mbox{bulk}}}$ is the $\gamma$-factor of the bulk moving towards the observer. The $\Gamma_{\tiny{\mbox{bulk}}}$-factor appears in eq.~(\ref{eq:flux}) because the dominant synchrotron radiation is emitted in the direction parallel to the motion of the bulk and it is confined within a cone whose angle is of the order $\pi/\Gamma^{2}_{\tiny{\mbox{bulk}}}$.  Substituting eq.~(\ref{eq:power}) into (\ref{eq:flux}), a straightforward evaluation yields
 \begin{eqnarray}
 \mathcal{F}&\sim& \left(1.6\times 10^{-35}\Gamma^{2}_{\tiny{\mbox{bulk}}}\right)
 \left[\frac{(\sin\delta)^{(p+1)/2}}{p+1}\Gamma\left(\frac{p}{4}+\frac{19}{12}\right)
 \Gamma\left(\frac{p}{4}-\frac{1}{12}\right)\right]\nonumber\\
 &&\times\left[1.9\times10^{-8}\left(\frac{\omega}{\mbox{s}^{-1}}\right)
 \left(\frac{\mbox{Gauss}}{|\Bbf|}\right)\right]^{-(p-1)/2}
 \left(\frac{C}{\mbox{m}^{-3}}\right)\left(\frac{B}{\mbox{Gauss}}\right)
 \left(\frac{V_{\tiny{\mbox{jet}}}}{\mbox{m}^{3}}\right)
 \left(\frac{\mbox{pc}}{D}\right)^{2}\mbox{Jy}.\nonumber\\
\end{eqnarray}

In neutron star mergers if we were to model a solar-mass neutrino jet to produce an observable signal through synchrotron radiation, these jets would have to be narrow to give the required neutrino number density. Lower energy neutrinos yield a larger neutrino density but in the absence of a solid model of neutrino jet production it is impossible to make a more concrete prediction. Another scenario of observability could be neutrino jets carrying a few to several hundred solar masses of energy. This could be realized for super-massive black holes. While the spectrum of the signal will depend on the strength of the magnetic field and the plasma through which a neutrino jet passes, its duration of a few seconds is the best quantitative probe to hunt for the physics discussed here. We are also assuming that CP is strongly violated so as to produce an excess of neutrinos or anti-neutrinos.

Quantitatively, a solar mass neutrino jet with a pulse length of one second and radius of one kilometer has $V_{\tiny{\mbox{jet}}}\approx 9.42\times 10^{14}\,\mbox{m}^{3}$ and $n_\alpha \approx 2.8 \times 10^{37} \mbox{cm}^{-3}$ for MeV neutrinos. If we now take $p=2.5$ which is a reasonable value for many astrophysical scenarios, $C=10^{3}\,\mbox{m}^{-3}$, $\omega=1.4\,\mbox{GHz}$ and $\Gamma_{\tiny{\mbox{bulk}}}=50$, the spectral flux density is
\begin{equation}
\mathcal{F}\sim (10^{-16}\,\mbox{Jy})\left(\frac{\mbox{pc}}{D}\right)^{2}\left(\frac{|\Bbf|}{\mbox{Gauss}}\right)^{7/4}.
\end{equation}
Therefore, if the neutrino masses are of the order indicated by eq.~(\ref{eq:mi}), the magnetic field is $|\Bbf|\sim 10^{-4}\,\mbox{Gauss}$. Taking the minimal observable spectral flux density to be of the order $10^{-6}\,\mbox{Jy}$, the source must be within a distance of $D\lesssim 10^{-9}\,\mbox{pc}$ or $D\lesssim 10^{-6}\,\mbox{pc}$ if we use the LSND suggested masses. This possibility is ruled out as there are clearly no sources within such distances since they are smaller than $1\,\mbox{AU}$.


The best chance to detect such pulses comes from well motivated models beyond the minimal extended SM where the magnetic moments can be as large as $10^{-12}\mu_{B}$~\cite{Giunti:2014ixa,Studenikin:2018vnp}. It then follows that for $n_\alpha \approx 2.8 \times 10^{37} \mbox{cm}^{-3}$, we would have $|\Bbf|_{\tiny{\mbox{max}}}\sim 10^6\,\mbox{Gauss}$ and
$
\mathcal{F}_{\tiny{\mbox{max}}}\sim (10^{-6}\,\mbox{Jy})(\mbox{pc}/D)^{2}
$
which is within the range of observability for $D\lesssim1\,\mbox{pc}$. A characteristic signature would be a short oscillating pulse of radiation that lasts for a few seconds.
 Realizations of these models would have a dramatic observable effect in discussed astrophysical situations. However, it seems unlikely that suitable sources can be found within a radius of 1 pc. Nevertheless, in such scenarios, a merger of neutron stars would leave behind a short pulse of synchrotron radiation with oscillating behavior lasting for a few seconds even if the neutrino jet in the merger points away from the neutrino detectors~\cite{PhysRevLett.119.161101}.  A typical neutron star has magnetic field of the order $|\Bbf_{n}|\sim10^{12}\,\mbox{Gauss}$ which is much larger than its counterpart produced by the neutrino jet. Therefore, in the presence of $\Bbf_{n}$, the contribution to synchrotron radiation emission from the neutrino jet would be negligible.
This means that to distinguish the pulse from synchrotron radiation produced by $\Bbf_{n}$, the jet must be sufficiently far away from the neutron star. 
 Observation of such a pulse would lend strong support to the indicated models of neutrino magnetic moments and solve the missing neutrino problem. On the other hand, the absence of such a pulse, when combined with detailed modeling of neutrino jets would provide new constraints on the neutrino magnetic moments and physics beyond the minimal extended SM.


\acknowledgments
C-YL is grateful to Debbijoy Bhattacharya, Kazuyuki Furuuchi, Noel Jonathan Jobu and Krishna Mohana for useful discussions.

\end{document}